\documentstyle[12pt]{article}
\title{Canonical Transformations in a Higher-Derivative Field Theory}
\author{ \sc{L. V. Belvedere, R. L. P. G. Amaral }\\
\small{\it Instituto de F\'{\i}sica - Universidade Federal Fluminense}\\
\small{\it Outeiro de S\~ao Jo\~ao Batista s/n, \,24020-005 Centro, Niter\'oi,
Rio de Janeiro - Brazil}\\
\\
\sc{N. A. Lemos}\thanks{On leave of absence from Departamento de
F\'{\i}sica, Universidade Federal Fluminense, Outeiro de S\~ao Jo\~ao
Batista s/n,\, 24020-005 Centro, Niter\'oi, RJ, Brazil.}\\
\small{\it Center for Theoretical Physics}\\
\small{\it Laboratory for Nuclear Science
and Department of Physics}\\
\small{\it Massachusetts Institute of Technology}\\
\small{\it Cambridge, MA 02139 - USA} }
\date{\today}
\begin{document}
\pagestyle{myheadings}
\baselineskip 18pt

\maketitle
\begin{abstract}
It has been suggested that the chiral symmetry can be implemented only in
classical Lagrangians containing higher covariant derivatives of odd order.
Contrary to this belief, it is shown that one can construct an exactly soluble
two-dimensional higher-derivative
fermionic quantum field theory
containing only derivatives of even order whose classical Lagrangian
exhibits chiral-gauge invariance.
The original
field solution is expressed in
terms of usual Dirac spinors through a canonical transformation, whose
generating function allows the determination of the new Hamiltonian. It is
emphasized that the original and  transformed Hamiltonians
are different because the mapping
from the old to the new canonical variables depends explicitly on time.
The violation of cluster decomposition is discussed and the general
Wightman functions satisfying the positive-definiteness condition are
obtained.
\\

\end{abstract}
\newpage
\centerline{\Large{\bf Introduction}}
\vspace{0.75cm}
\noindent In the last few years several authors have turned their attention to
the
study of higher-derivative  field theories. The reader is referred to [1-3]
for the motivation of such investigations, as well
as for a brief sketch of the historical development  of the subject.
It is the purpose of this paper to show that certain statements
concerning higher-derivative field theories
with chiral symmetry are based on hasty arguments, and to make the
discussion concrete a two-dimensional higher-derivative free model of
even order and its local gauge invariant generalization
(higher-derivative Schwinger model of even-order) are considered.
We also address ourselves to  some unexplored
aspects of two-dimensional quantum field theories with higher-derivative
couplings.

One occasionally finds in the literature statements to the effect that
the chiral symmetry restricts ``the number
and the kind'' of covariant derivatives in the fermionic Lagrangian, so that,
for instance, only the appearence of an odd number of covariant derivatives of
the fermion fields is compatible with invariance (at the classical level) of
the fermionic Lagrangian under chiral-gauge transformations [3]. Here
we show that this is not the case by explicitly constructing a
higher-derivative model of even order whose classical Lagrangian exhibits
chiral-gauge invariance.

The local operator solution of this higher-derivative model is obtained in
terms of a product of local spacetime functions and usual ``bona fide'' Dirac
spinors, thus defining a mapping between two sets of field operators. It is
pointed out that, being explicitly time dependent, such a mapping leads
to a new Hamiltonian $ H^{\,\prime } $ that differs from the original
Hamiltonian $ H $. The canonical character of the transformation is stressed
and the generating function involving old and new variables is obtained. The
same analysis can be extended to higher-derivative models of odd order.

The theory involves an indefinite-metric ``Hilbert space''
and cluster decomposition is violated. It is shown, however,
that the general Wightman functions for the canonical free and massless
Dirac fields satisfying the positive-definiteness condition can be recovered
by considering correlations between the original field variables and
their conjugate momenta.
\vspace{0.75cm}

\centerline{\Large{\bf 1  Free Model}}

\vspace{0.75cm}

Attention has
been called recently [2,3] to Lorentz invariant
field models in two-dimensional spacetime defined by the
Lagrangian density
\\
$$ {\cal L}_o\,=\,i\,\bar {\xi }\,(\,\partial\!\!\!/ \,
\partial\!\!\!/ ^{\,\dagger }\,)^{\,N}\partial\!\!\!/\,\xi \,\,\,,\eqno (1.1)$$
\newpage
\noindent whose local gauge-invariant generalization can be exactly
solved\footnote{The conventions used
are : \,$ \xi \,=\,
(\,\xi _{(1)} \,,\,\xi _{(2)})^{\,T} $\,\,,\,\,$ \epsilon
^{\,0\,1}\,=\,g^{\,0\,0}\,
=\,-\,g^{\,1\,1}\,=\,1 $\,\,,
\newline
\vspace{0.1cm}
\centerline{$ \gamma ^{\,5}\,=\,-\,\gamma ^{\,0}
\gamma ^{\,1}\,\,,\,\,\gamma ^{\,\mu }\gamma ^{\,5}\,=\,\epsilon ^{\,\mu \,\nu}
\,\gamma _{\nu } $}
\newline
\vspace{0.1cm}
\centerline{$ \gamma ^{\,0}\,=\,\pmatrix {0&1 \cr 1&0} $\,\,,\,\,
$ \gamma ^{\,1}\,=\,\pmatrix {0&1 \cr -1&0} $ \,\,,\,\, $ \gamma ^{\,5}\,=\,
\pmatrix {1&0 \cr 0&{-1}} $}}.

With the purpose of introducing the extension to models with an even number
of field derivatives, let us consider the Lorentz transformation (L\,T)
properties in
two dimensions. In this case, the behavior of fields under a L\,T is better
analyzed using the light-cone
variables : $ x^{\,\pm }\,=\,x^{\,0}\,\pm \,x^{\,1} $. Under a L\,T these
variables transform according
with \,\,$ x^+ \rightarrow \lambda \,x^+ $\,\,,\,\,$ x^- \rightarrow \lambda
^{-1}\,
x^- $\,\,, such
that \,\,$\partial _+ \rightarrow \lambda ^{-1} \partial _+ $\,\, and
\,\,$ \partial _- \rightarrow \lambda  \partial _- $\,\,, where  $\lambda \in
(0 , \infty)$.
In terms of spinor
components $ \xi _{(\alpha )} $, the kinetic term (1.1) can be written as
\\
$$ {\cal L}_o\,=\,i\,\xi _{(1)}^{\,*}\,\partial_- ^{\,2N + 1}\,\xi  _{(1)}\,+
\,
i\,\xi  _{(2)}^{\,*}\,\partial_+^{\,2N + 1}\,\xi _{(2)}
\,\,\,. \eqno (1.2)$$
\\
Thus, under a L\,T the \,\,upper \,\,and \,\,lower \,\,spinor
\,\,components\,\,
obey \,\, the \,\, transformation \,\, law \hfill
\newline $ \xi  _{(\alpha )}\,\rightarrow \lambda ^{(N+{1\over 2})\gamma
^{\,5}_{\alpha \alpha }}\,
\xi  _{(\alpha )} $.

Taking these considerations into account, the extension to the case with an
even number of Fermi fields derivatives can be introduced through the
following (Hermitian) Lagrangian density
\\
$$ {\cal L}_o\,=\,\bar {\xi }\,\gamma ^{\,0}\,(\,i\,/\!\!\!\partial ^{\,\dagger
}\,
i\,/\!\!\!\partial \,) ^{\,N}\,\xi
\,\,\,,\eqno (1.3) $$
\\
where the imaginary factor $ i $ was inserted for future convenience. In this
case, the Lorentz transformation properties of the spinor components
are  $ \xi _{(\alpha )}\,\rightarrow \lambda ^{\,-\,N \gamma^{\,5}_{\alpha
\alpha}}\,
\xi _{(\alpha )} $.

Let us now perform the quantization of the free model defined by the Lagrangian
(1.3). For the sake of simplicity we consider the upper component only.
Proceeding in the spirit of Ref. [2], we consider the configuration space
generated by
\\
$$ \xi _{(1)}^{\,n}\,=\,(\partial _-)^{\,n}\,\xi _{(1)} \,\,\,\,,\,\,\,\,
n\,=\,0,1,..,2N-1 \,\,\,\,, \eqno (1.4)$$
\\
so that the associated momenta are given by
\\
$$\pi ^{\,n}_{\xi }\,=\,(-1)^{\,(N-1-n)}\,(\partial _-)^{\,(2N-1-n)}\,\xi
_{(1)}^{\,*}
\,\,\,\,. \eqno (1.5) $$
\\
Since the system under consideration exhibits constraints, the canonical
quantization must be per-
\newpage
\noindent formed using the Dirac method, which leads to the
equal-time anticommutators
\\
$$ \Bigl \{\,\partial _- ^{\,p}\,\xi _{(1)}(x)\,,\,\partial _- ^{\,q}\,
\xi _{(1)}^{\,*}
(y)\,\Bigr \}_{ET}\,=\,i\,(-1)^{N-p-1}\,\delta _{q+p\,,\,2N-1}\,\delta
(x^1-y^1)
\,\,\,\,. \eqno (1.6) $$
\\
In momentum space, the solution of the equation of motion for $\xi _{(1)}$
is a linear combination of derivatives
of $ \delta (\,k^{\,2}\,) $ up to the order $(N-1)$. The Fourier representation
for the operator solution of the model which leads to a local field operator
is given by ($ k\,\equiv \,k_+ $)
\\
$$ \xi _{(1)}(x)\,=\,\frac{ m^{\,{1\over 2} -N}}{\sqrt {2\,\pi }}\,\int
_{-\infty } ^{+\infty } d\,
k\,e^{\,-\,ik\,(\,x^{\,0}\,+\,x^{\,1}\,)}\,
\sum _{p = 0} ^{2N-1}\, {(-\,imx^-)^{\,p}\over p\,!}\,\tilde \xi
_{(1)}^{\,p}(k)
\,\,\,\,.\eqno (1.7) $$
\\
In the above expression  the finite arbitrary mass scale $ m $ is
introduced\footnote{It should be remarked that there is another solution, in
which the arbitrary dimensional parameter $ m $ is not introduced, but it leads
to a
non-local field operator. In order to circumvent this
problem and obtain a solution in terms of usual fermionic field operators, the
parameter $ m $ must be introduced. In this case, the previously
referred Lorentz transformation properties of the spinor
components $ \xi _{(\alpha )} $ are implemented if we perform the LT
combined with the redefinition $ m \,\rightarrow \,\lambda m $.} in order to
ensure the usual
dimension for the spinor component
fields  $ \tilde \xi _{(1)}^{\,p} (k) $, which satisfy
\\
$$\Bigl \{\,\tilde \xi _{(1)}^{\,p}\,(k)\,,\,\tilde \xi _{(1)}^{\,q\,*}
(k^{\,\prime })\,
\Bigr \}\,=\,\delta_{q+p\,,\,2N-1}\,\delta (k-k^{\,\prime })\,\,\,\,.\eqno
(1.8)$$
\\

The anticommutation relations of the mode
expansion operators can be diagonalized via
the linear transformation
\\
$$ \tilde \psi _{(1)}^{\,p}(k)\,=\,{1\over \sqrt {2}}\,\bigl (\,
\tilde \xi _{(1)}^{\,p-1}(k)\,+\,
\tilde \xi _{(1)}^{\,2N-p}(k)\,\bigr )\,\,\,\,,\eqno (1.9a) $$
\\
$$ \tilde \psi _{(1)}^{\,N+p}(k)\,=\,{1\over \sqrt {2}}\,\bigl
(\,\tilde \xi _{(1)}^{\,p-1}(k)\,-\,
\tilde \xi _{(1)}^{\,2N-p}(k)\,\bigr )\,\,\,\,,\eqno (1.9b) $$
\\
with $ p\,=\,1,2,..,N $. Now a set of $ 2N $ usual free Dirac
spinors in coordinate space $ {\psi}^{\, j} $ can be

\noindent introduced. Defining their upper components by
\\
$$ \psi _{(1)}^{\,p}(x)\,=\,{1\over \sqrt{2\pi }}\,\int d\,k\,e^{\,-\,k\,\cdot
\,
(x^{\,0}\,+\,x^{\,1}\,)}\,\tilde \psi _{(1)}^{\,p}\,(k)\,\,\,\,,\eqno (1.10) $$
\\
we can write
\\
$$\xi _{(1)}(x)\,=\,\sum _{j=1} ^{2N}\,f_j (x^- )\,\psi _{(1)}^{\,j}\,(x^+ )
\,\,\,\,. \eqno (1.11) $$
\\
The Dirac fields $ \psi _{(1)}^{\,j} (x^+ ) $ are quantized with positive
(negative) metric for $ j\leq N\,(\,j>N\,) $. The factors $ f_j (x^- ) $ are
given by
\\
$$ f_j (x^- )\,=\,{m^{{1\over 2}-N} \over \sqrt {2} }\,\Bigl [\,
 {(-\,imx^- )^{\,j\,-\,1} \over (j\,-\,1)\,! }
\,+\,{(-\,imx^- )^{\,2N\,-\,j} \over (2N\,-\,j)\,! }\,\Bigr ]\,\,\,\,,
\eqno (1.12a) $$
\\
$$ f_{j\,+\,N}(x^- )\,=\,(\,-\,1\,)^{\,j\,+\,1}\, f_{j}^{\ast }(x^- )\,\,\,\,,
\eqno (1.12b) $$
\\
where $ j\,\leq \,N $. The expression for the lower
component $ \xi _{(2)}(x) $ is obtained interchanging the light-cone
variables $ x^+\,\leftrightarrow \,x^- $. In order to ensure the correct
Lorentz
transformation properties of the lower spinor component, a new dimensional
parameter $ \tilde m $ must be introduced such that under a Lorentz
transformation it is redefined
as $ \tilde m\,\rightarrow \,\lambda ^{\,-\,1}\,\tilde m $. In this way,
the higher-order ``spinor field'' can be expressed as a linear
combination of usual Dirac fields.

The anticommutator between $ \xi_{(\alpha)} $ ($\alpha = 1,2$) fields is given
by
\\
$$ \Bigl \{\,\xi_{(\alpha)}^{\,\ast }(x)\,,\,\xi_{(\alpha)}(y)\,\Bigr \}\,=\,
\frac{(-\,i)^{\,2N-1}}{(2N-1)!}\,
\Bigl (\,x^{\,\mp}\,-\,y^{\,\mp}\,\Bigr )^{2N-1}\,S\,(\,x^{\,\pm}\,-\,y^{\,\pm}
\,)  \eqno(1.13)$$
\\
where $ S(x) $ is the anticommutator of two usual free and massless Dirac
fields. Although the $ \xi $ field does not represent a genuine spinor field
operator, the
corresponding spinorial nature is carried by the
infrafermion $ \psi ^{\,j} $ field operators, which ensure the correct
micro-causality requirements. In this sense, the general quantum field features
are implemented by the infrafermion operators.

As expected from a theory with an indefinite-metric ``Hilbert space'', the
cluster
decomposition is violated. As a consequence of (1.13), for $ \alpha = 1 $ (2)
the $ 2p $-point
functions of the $ \xi_{(\alpha)} $ fields violate the cluster decomposition in
the light-cone variable $ x^{\,-} $ $ (x^{\,+}) $. Nevertheless, the general
Wightman functions generated by the canonical free and massless Dirac spinor
components $ \psi^{\,o} _{(\alpha)}(x^{\,\pm}) $ satisfying the
positive-definiteness condition can be recovered considering
arbitrary correlations between the configuration space
variable $ \xi^{\,\ell}(x) $ and the associated
momenta $ \pi^{\,\ell}_{\xi}(x) $ defined respectively by Eqs.(1.4) and
(1.5). Thus, for the $ 2p $-point function with
fixed  $ \ell = 0,...,2N - 1 $, we get
\\
$$ \langle\,0\,\vert\,\prod_{j=1}^{p}\,\pi^{\,\ell}_{\xi_{(\alpha )}}(x_j)\,
\prod _{k=1}^{p}\,\xi^{\,\ell}_{(\alpha)}(y_k)\,\,\vert\,0\,\rangle \,
=\, \langle\,0\,\vert\,\prod _{j=1}^{p}\,(\,
\partial_{\mp})^{\,2N-1-\ell }
\xi^{\,\ast}_{(\alpha)}(x_j)\,\prod
_{k=1}^{p}\,(\partial_{\mp})^{\ell }\xi_{(\alpha)}(y_k)\,
\vert\,0\,\rangle \,\equiv $$
\\
$$ \equiv \,c\,\langle \,0\,\vert \,\prod _{j=1}^{p}\,
\psi^{o\,\ast}_{(\alpha)}(x^{\,\pm}_j)\,\prod _{k=1}^{p}\,
\psi^{\,o}_{(\alpha)}(y^{\,\pm}_k)\,\vert\,0\,\rangle \,=\,
c\,\frac{{\prod\atop{i<j}}(x_i - x_j){\prod\atop{k<m}}(y_k - y_m)}
{{\prod\atop{i\neq k}}(x_i - y_k)}\,\,. \eqno(1.14)$$
\\
where $ c $ is a constant.

The bosonization of the free theory can be carried out
using the stardard bosonization formulae, leading to $ N $ independent
$ \phi _j $ scalar fields associated to the Dirac spinor $ \psi ^j $. Similarly
to what is done in
Refs.[2,4], we introduce the following decomposition for the $ 2N $ independent
scalar fields :
\\
$$ \phi _j\,=\,\frac{1}{\sqrt {2N}}\,\Phi \,+\,\sum _{i_D\,=\,1}^{2N\,-\,1}\,
\lambda _{j\,j}^{\,i_D}\,\varphi ^{\,i_D}\,\,\,\,, \eqno (1.15)$$
\\
where $ \lambda _{j\,j}^{\,i_D} $ are the diagonal matrices of the
$ {\cal S}{\cal U}(2N) $. The $ \Phi $ field acts as the potential for
$ {\cal U}(1) $ and chiral-${\cal U}(1) $ conserved currents. Using this
decomposition in the usual bosonized expression for the free Dirac
fermion operator, the $ {\cal U}(1) $ and chiral-${\cal U}(1) $ dependence
factorize and we get
\\
$$ \xi _{(1)}(x)\,=\,S_{(1)}^{\Phi }(x^+)\sum _{j\,=\,1}^{2N}\,
f_j(x_-)\,{\hat \psi }_{(1)}^{\,j}(x^+)\,\,\,\,,\eqno (1.16) $$
\\
where the `` soliton '' operator $ S^{\Phi } $  and the $ {\cal U}(1) $
screened Dirac infrafermions $ \hat \psi ^{\,j} $ are given by [4]
\\
$$ S_{(1)}^{\Phi }(x^+)\,=\,:\,\exp
\,\Biggl\{\,2\,i\,\sqrt{\frac{\pi}{2N}}\,\Phi (x^+)
\Biggr \}\,:\,\,\,\,,\eqno (1.17a) $$
\\
$$\hat \psi _{(1)}^{\,j}(x^+)\,=\,\Bigl (\,\frac{\mu}{2\,\pi }\,
{\Bigr )}^{\,\frac{1}{2}}\,:\,\exp \,\Biggl \{\,2\,i\,\sqrt \pi\,
\sum _{i_D\,=\,1}^{2N\,-\,1}\,\lambda _{j\,j}^{\,i_D}\,\varphi ^{\,i_D}(x^+)
\,\Biggr \}\,:\,\,\,\,.\eqno (1.17b) $$
\\
In the expressions above we have suppressed the corresponding Klein factors
needed to ensure the correct anticommutation relations [5]. As
stressed in [2], these Klein factors are also  introduced in order to implement
the indefinite metric Hilbert space.
\vspace{0.75cm}

\centerline{\Large{\bf 2 Canonical Transformation }}

\vspace{0.75cm}

In this section we shall obtain the Hamiltonian $ H $ for the model defined
by the Lagrangian density (1.3), which evolves the $ \xi $ field, and show
that it is related by a canonical transformation to the
Hamiltonian $ H^{\prime } $ that describes the time evolution of the
usual Fermi field $ \psi $. For the sake of simplicity we consider the
upper component only.

To begin with we use a generalized Legendre transformation of the
Lagrangian density (1.3), such that the Hamiltonian density $ {\cal H} $ can
be written as
\\
$$ {\cal H}(x) = \sum _{n=0} ^{2N-1}\,\pi^{\,n}_{\xi}(x)\,
\dot \xi ^{\,n}_{(1)}(x)\,-\,\xi _{(1)}^*(x) (i\partial _- )^{2N}\xi _{(1)} (x)
\,\,\,\,.\eqno (2.1a)$$
\\
Using the definitions (1.4) for the configuration space field variables
and (1.5) for the corresponding associated momenta, we obtain the Hamiltonian
density (2.1a) in terms of the phase-space variables as being
\\
$$ {\cal H}(x)\,=\,\sum _{n=0} ^{2N-2}\,(-1)^{N-1-n} \xi
_{(1)}^{\ast\,{2N-1-n}}
\,\xi _{(1)}^{\,n+1}(x)\,+\,\sum_{n=0}^{2N-1}\,(-1)^{N-1-n}\,
\xi _{(1)}^{*\,2N-1-n}(x)\,\partial _1\,\xi _{(1)}^{\,n}(x)\,\,\,\,.
\eqno (2.1b)$$
\\
The corresponding  Hamiltonian $ H $ evolves the field $\xi _{(1)}(x)$.
Since the expression of the original configuration space variables
in terms of the $ 2N $ usual Dirac fields $ \psi ^{\,j} $ is
explicitly time dependent, the
Hamiltonian obtained from (2.1) does not describe the time evolution of the
latter
fields. Nevertheless let us show that the Hamiltonian $ H^{\,\prime } $ that
describes the time evolution of the Dirac fields can be obtained
from $ H $ by canonical methods.

To this end, we consider Eq.(1.11) as defining a mapping from the set of
old variables $ \xi ^{\,p}_{(1)} $, given by Eq.(1.4), to the set of new
variables $ \psi ^{\,p}_{(1)} $,
\\
$$ \{ \xi ^{\,n}_{(1)}(x) \} \rightarrow \{ \psi ^{\,p}_{(1)}(x) \}\,\,\,\,,
\eqno (2.2) $$
\\
Taking into account the expressions for all the original phase space
variables given by (1.11), we have the set of transformation equations:
\\
$$ \xi ^{\,n}_{(1)}(x)\,=\,\sum _{j=1}^{2N}\,\Bigl [\,\partial _-^{\,n}
f_j(x^-)\,\Bigr ]\psi _{(1)}^{\,j}(x)\,\,\,\,.\eqno (2.3)$$
\\
After this point transformation, the new momenta are obtained computing
the variation of the action around a solution of the equation of motion:
\\
$$\delta S = \int _{t_1} ^{t_2} d^{\,2}x\,\xi _{(1)}^* (x) (i\partial _-)^{2N}
\delta \xi _{(1)}(x)\,\,\,\,.\eqno (2.4) $$
\\
Considering vanishing variations at the initial time $t_1$, we get
\\
$$\delta S = i \sum _{j=1} ^N\,\Bigl [\,\psi _{(1)} ^{\ast j}\,(x)\,\delta \,
\psi _{(1)}^{\,j}
\,(x)\,-\,\psi _{(1)}^{\ast (j+N)}\,(x)\,\delta \,\psi _{(1)}^{\,j+N}(x)\,
\Bigr ]_{t\,=\,t_2}\,\,\,\,.
\eqno (2.5)$$
\\
and we can identify $ i\,\psi _{(1)}^{\ast j} $ as the canonical conjugate
momenta
of the $ \psi _{(1)}^{\,j} $ fields. The anticommutators for the Dirac fields
are
thus reobtained without resort to the anticommutation relations of the
original field solution.

In order to obtain the expression of the new Hamiltonian, let us consider the
generating function for the point transformation (2.3). In classical
mechanics the generating function of a point transformation from the set of
conjugate variables $ \{ p_i,q_i\} $ to the set $ \{ P_i,Q_i\} $ can be
written as
\\
$$ \Omega (P , q, t)\,=\,\sum _i P_i \, Q_i(q,t)\,\,\,\,.\eqno (2.6) $$
\\
The transformation equations can be recovered  by means of
\\
$$ Q_i\,=\,\frac{\partial \Omega }{\partial P_i } \,\,\,\,\,\,,\,\,\,\,\,\,
p_i\,=\,\frac{\partial \Omega }{\partial q_i}\,\,\,\,,\eqno (2.7) $$
\\
while the expression for the new Hamiltonian function is given by
\\
$$ {\cal H}^{\,\prime }\,=\,{\cal H}\,+\,\partial _o \,\Omega \,\,\,\,.\eqno
(2.8)$$
\\

The generalization of this formalism to higher-derivative field
theory can be
achieved by considering the ``generating function''\footnote{ Strictly speaking
one should consider a generating
functional obtained by integrating $ \Omega _{(1)}(x) $ over space.
In the peculiar case dealt with here this is not necesssary, and
the formalism  with generating function works quite satisfactorily.}
of the transformation from the
{\it new} $ \psi ^{\,j} $ variables to the {\it old } $ \xi ^{\,n} $
variables as the sum of the old momenta multiplied by the corresponding old
variables written in terms of the new ones, that is,
\\
$$ \Omega _{(1)}(x)\,=\,\sum _{n=0}^{2N-1}\,\Biggl \{\,(-1)^{N-1-n}\,
\xi ^{\,*(2N-1-n)}_{(1)}(x)\,\sum _{j=1}^{2N}
\,\Bigl [\,\partial _-^{\,n} f_j(x^-)\,\Bigr ]\,\psi _{(1)}^{\,j}(x^+)\,\Biggr
\}
\,\,\,\,.\eqno (2.9)$$
\\
This generating function  allows one to obtain the transformation (2.3)
through
\\
$$ \xi _{(1)}^{\,n} (x)\,=\,(-1)^{\,N-1-n}\,
\frac{\partial \Omega (x)}{\partial {\xi }^{\,\ast \,2N-n-1}_{(1)}(x)}\,\,\,\,,
\eqno (2.10a)$$
\\
$$ i(-1)^{\,\ell }\,\psi _{(1)}^{\,j\ast }\,=\,
\frac{\partial \Omega (x)}{\partial \psi _{(1)}^{\,j}(x)}\,\,\,\,,\eqno (2.10b)
$$
\\
where $ \ell \,=\,0\,(1) $, for $ j\,\leq N \,(\,j\,>\,N\,) $. The
new Hamiltoninan density $ {\cal H}^{\,\prime } $ differing
from $ {\cal H} $ by $ \,\partial _o \Omega _{(1)}(x) $  is given by
\\
$${\cal H}^{\,\prime }\,=\,\sum _{j=1}^N\,\Bigl [\,i\,\psi _{(1)}^{\,*j}(x)\,
\partial _1\,\psi _{(1)}^{\,j}(x)\,-\,
i\,\psi _{(1)}^{\,*j+N} (x)\partial _1\psi _{(1)}^{\,j+N}(x)\,\Bigr ]\,\,\,\,,
\eqno (2.11)$$
\\
which is the correct Hamiltonian density associated with $ N $
independent free Dirac spinor components $ \psi _{(1)}^{\,j}(x) $, and the one
to be bosonized.

Analogous expressions are obtained for the lower component $ \psi
^{\,j}_{(2)}(x) $.
\vspace{0.75cm}

\centerline{\Large{\bf 3 Local Gauge Invariance}}

\vspace{0.75cm}

The Schwinger model with derivatives of order $ 2\,N $ is
defined by the Lagrangian density
\\
$$ {\cal L}\,=\,-\,\frac{1}{4}\,(\,{\cal F}_{\mu \,\nu }\,)^{\,2}\,+\,
\bar \Psi \,{\triangle \!\!\!\!/}^{\,2\,N}\,\Psi \,\,\,\,,\eqno (3.1)$$
\\
where $ {\triangle \!\!\!\!/}^{\,2\,N} $ is the covariant derivative of
order $ 2\,N $ defined by
\\
$$ {\triangle \!\!\!\!/}^{\,2\,N}\,=\,\gamma ^{\,0}\,(\,i\,
/\!\!\!\!{\cal D}^{\,\dagger }\,i\,/\!\!\!\!{\cal D}\,)^{\,N}\,\,\,\,,
\eqno (3.2)$$
\\
where the usual covariant derivative is  $  /\!\!\!\!{\cal D}\,=\,\gamma
^{\,\mu }\,
(\,\partial _{\mu }\,-\,i\,e\,{\cal A}_{\mu }\,) $.
\\

The quantization of the model can be performed following the same steps
outlined
in the free-field case. The configuration space is now generated by
\\
$$ \Psi ^{n}\,=\,{\cal D}_{\pm }^{\,n}\,\Psi \,\,\,\,\,\,\,\,
n\,=\,0,1,2,\dots ,2N\,-\,1\,\,\,\,,\eqno (3.3) $$
\\
and the associated momenta
\\
$$ \pi _{\Psi }^{\,n}\,=\,(-\,1)^{\,(N-1-n)}\,\Bigl (\,{\cal D}^{\,2N-1-n}\,
\Psi \,\Bigr )^{\,\ast }\,\,\,\,,\eqno (3.4) $$
\\
which satisfy canonical anticommutation relations. Since a chiral operator
gauge
transformation acting on the free field decouples the gauge field, similarly to
what is done in Ref.[2], the configuration space variables can be written as
\\
$$ \Psi ^{n}(x)\,=\,:\,e^{\,i\,\sqrt{\frac{\pi }{2N}}\,\gamma ^{\,5}\,[\,
\tilde \Sigma (x)\,+\,\tilde \eta (x)\,]}\,:\,\xi ^{n}(x)
\,\,\,\, , \eqno (3.5)$$
\\
$$ \pi _{\Psi }^{\,n}(x)\,=\,:\,e^{\,-\,i\,\sqrt{\frac{\pi }{2N}}\,\gamma
^{\,5}\,[\,
\tilde \Sigma (x)\,+\,\tilde \eta (x)\,]}\,:\,\pi _{\xi } ^{n}(x)
\,\,\,\, . \eqno (3.6)$$
\\

The operator solution for the
even order case is then given by
\\
$$ \Psi_{(\alpha )} (x)\,=\,:\,\exp \,\Biggl \{\,i\,\Bigl (\,\frac{\pi
}{2\,N}\,
\Bigr )^{\,\frac{1}{2}}\,\gamma_{\alpha \alpha}^5\,[\tilde \Sigma (x)\,+\,
\tilde \eta (x)\,]\,\Biggr \}\,:\,\sum_{j=1}^{2N}\,f_j(x^{\,\mp})\,
\psi ^{\,j}(x^{\,\pm})\,\,\,\, . \eqno(3.7) $$
\\
Performing an operator gauge transformation
\\
$$ \hat \Psi_{(\alpha )} (x)\,=\,:\,e^{\,i\,\sqrt{\frac{\pi}{2N}}\,\eta
(x)}\,:\,
\Psi_{(\alpha )} (x) \,\,\,,\eqno(3.8) $$
\\
we obtain the ``physical fermion field operator''
\\
$$ \hat \Psi_{(\alpha )} (x)\,=\,:\,\exp \,\Biggl \{\,i\,\Bigl (\,\frac{\pi
}{2\,N}\,
\Bigr )^{\,\frac{1}{2}}\,\gamma_{\alpha \alpha}\,\tilde \Sigma (x)\,
\Biggr \}\,:\,\sum_{j=1}^{2N}\,f_j(x^{\,\mp})\,
\hat \psi ^{\,j}(x^{\,\pm})\,\sigma \,\,\,\, , \eqno(3.9) $$
\\
where $ \sigma $ is a spurious operator with scale dimension of zero value and
given in terms of free and massless scalar fields by
\\
$$ \sigma \,=\,\exp \,\Biggl \{\,i\,\Bigl (\,\frac{\pi }{2\,N}\,
{\Bigr )}^{\,\frac{1}{2}}\,\{\,\gamma ^{\,5}\,[\,\tilde \Phi (x)\,+\,
\tilde \eta (x)\,]\,+\, [\,\Phi (x)\,+\,\eta (x)\,]\,\}\,\Biggr \}\,\,\,\,,
\eqno (3.10)  $$
\\
where the $ \eta $ field is quantized with negative metric. The gauge field is
given by
\\
$$ \hat {\cal A}_{\mu }(x)\,=\,-\,\frac{1}{e}\,\sqrt \frac{\pi }{2\,N}\,
\varepsilon _{\mu \,\nu }\partial ^{\,\nu }\,\tilde \Sigma (x)\,\,\,\,,
\eqno (3.11) $$
\\
with $ \tilde \Sigma $ satisfying the equation of motion
\\
$$ \Biggl [\,\Box \,+\,\frac{2\,N\,e^{\,2}}{\pi }\,\Biggr ]\,
\tilde \Sigma (x)\,=\,0 \,\,\,\,,\eqno (3.12) $$
\\
and the anomalous divergence of the axial current is then given by
\\
$$ \partial ^{\,\mu }\,\Im ^{\,5}_{\mu }\,=\,-\,\frac{N\,e}{\pi }\,
\varepsilon _{\mu \,\nu }\,{\cal F}^{\,\mu \,\nu }\,\,\,\,. \eqno (3.13) $$
\\

The generalized fermion Green functions are obtained considering
correlators like
\\
$$ \langle \,0\,\vert \,\prod _{j=1}^{p}\,
\Bigl ({\cal D}_{\mp}^{\,2N-1-\ell}\,\Psi_{(\alpha)}
(x_j)\,\Bigr )^{\ast}\,\prod _{k=1}^{p}\,
{\cal D}_{\mp}^{\ell}\Psi _{(\alpha)}(y_k)
\,\vert \,0\,\rangle \,=\,
\langle \,0\,\vert \,\prod _{j=1}^{p}\,\psi^{o\,\ast}_{(\alpha)}(x^{\,\pm}_j)
\,\prod _{k=1}^{p}\,\psi^{\,o}_{(\alpha)}(y^{\,\pm}_k)\,
\vert\,0\,\rangle\,\times $$
\\
$$\times\,\langle \,0\,\vert\,\prod_{j=1}^p\,:\,
e^{\,-\,i\,\sqrt {\frac{\pi }{2N}}\,\gamma^{\,5}_{\alpha \alpha}\,[\,
\tilde \Sigma (x_j)\,+\,\tilde \eta (x_j)\,]}\,:\,\prod_{k=1}^p\,
:\,e^{\,i\,\sqrt {\frac{\pi }{2N}}\,\gamma^{\,5}_{\alpha \alpha}\,[\,
\tilde \Sigma (y_k)\,+\,\tilde \eta (y_k)\,]}\,:\,\vert \,0\,\rangle
\,\,\,\, . \eqno(3.14)$$

The Wightman functions related to the physical fermion field
operators $ \hat \Psi $, are given by
\\
$$ \langle \,0\,\vert \,\prod _{j=1}^{p}\,
\Bigl ({\cal D}_{\mp}^{\,2N-1-\ell}\,
\hat \Psi_{(\alpha)}(x_j)\,\Bigr )^{\ast}\,\prod _{k=1}^{p}\,
{\cal D}_{\mp}^{\ell}\hat \Psi _{(\alpha)}(y_k)\,
\vert \,0\,\rangle \,=\, \langle \,0\,\vert \,
\prod _{j=1}^{p}\,\hat\psi^{\,\ast}_{(\alpha)}(x^{\,\pm}_j)
\,\prod _{k=1}^{p}\,\hat\psi_{(\alpha)}(y^{\,\pm}_k)\,
\vert\,0\,\rangle\,\times $$
\\
$$\times\,\langle \,0\,\vert\,\prod_{j=1}^p\,:\,
e^{\,-\,i\,\sqrt {\frac{\pi }{2N}}\,\gamma^{\,5}_{\alpha \alpha}\,
\tilde \Sigma (x_j)\,}\,:\,\prod_{k=1}^p\,
:\,e^{\,i\,\sqrt {\frac{\pi }{2N}}\,\gamma^{\,5}_{\alpha \alpha}\,
\tilde \Sigma (y_k)\,}\,:\,\vert \,0\,\rangle \,\,\,\,, \eqno(3.15)$$
\\
in which we have used the fact that
\\
$$ \langle \,0\,\vert\,\prod_{j=1}^p\,\sigma^{\ast}_{\alpha
}(x_j)\,\prod_{k=1}^p\,
\sigma_{\alpha}\,(y_k)\,\vert \,0\,\rangle \,=\,1 \,\,\,\, . \eqno(3.16)$$
\\
The Wightman functions (3.15) correspond to those of a subspace of the
Schwinger model with $ 2N $ flavored fermions introduced in Ref.[4].

\vspace{0.75cm}

\centerline{\Large{\bf 4 Final Comments}}

\vspace{0.75cm}

By way of conclusion, let us summarize our findings.
We have further extended the class of models amenable to exact solution
studied in [2] by considering Lagrangians involving higher
derivatives of even order.
These Lagrangians are immediately seen to be Lorentz invariant by
using light-cone variables. The resulting unusual behavior
of the fields in these generalized Schwinger models has been stressed. As a
final result the bosonization of models with
Lagrangians of the form
\\
$$ L = \xi ^{\dagger } ( i \gamma ^0 \partial \!\!\!/ )^N \xi \,\,\,\,,
\,\,\,\,N\,=\,1,2,3,...$$
\\
has been obtained as well  as of their gauged
versions. Special care, irrespective of $ N $ being even or odd, is needed when
the bosonization of expressions as the Lagrangian or the Hamiltonian
is undertaken. Indeed,
by considering the solution of the model as a point transformation of
fields the generating function for this transformation has been
obtained and therefrom the new Hamiltonian.
Owing to explicit spacetime dependence of the original
fields in terms of the ultimately bosonized internal fields $\psi^{\,j}$ ,
the bosonized Hamiltonians of the latter are not obtained by just
bosonizing the original Hamiltonians expressed in terms
of the new fields.


\newpage
\noindent{\bf  Acknowledgment} : The authors express their thanks to Conselho
Nacional de Desenvolvimento
Cient\'{\i}fico e Tecnol\'ogico (CNPq), Brazil, for partial financial support.
\vspace{1.0cm}

\centerline{\bf References}
\begin{description}
\item[1 -] C. Batlle, J. Gomis, J. M. Pons and N. Roman-Roy, J.Phys.
           {\bf A21}, 2963(1988);
\item      V. V. Nesterenko, {\it{ibid}} {\bf A22}, 1673(1989);
\item      C. A. P. Galv\~ao and N. A. Lemos, J. Math. Phys. {\bf 29},
1588(1988);
\item      J. Barcelos-Neto and N. R. F. Braga, Acta Phys. Polonica
{\bf 20},
           205(1989);
\item      J. Barcelos-Neto and C. P. Natividade, Z. Phys. {\bf C49},
511(1991);
\item      {\it{ibid}} {\bf C51}, 313(1991);
\item      J. J. Giambiagi, Nuovo Cimento {\bf 104 A}, 1841 (1991);
\item      C. G. Bolini and J. J. Giambiagi, J. Math. Phys. {\bf 34}, 610
(1993)
\item[2 -] R. L. P. G. do Amaral, L. V. Belvedere, N. A. Lemos and
           C. P. Natividade, Phys. Rev. {\bf D47}, 3443(1993)
\item[3 -] J. Barcelos-Neto and C. P. Natividade, Z.  Phys. C {\bf 49}, 511
           (1991)
\item[4 -] L. V. Belvedere, J. A. Swieca, K. D. Rothe and B. Schroer, Nucl.
           Phys. {\bf B 153}, 112 (1979);
\item       L. V. Belvedere, Nucl. Phys. {\bf B 276}, 197 (1986).
\item[5 -] M. B. Halpern, Phys. Rev. {\bf D 12}, 1684 (1975); {\bf 13}, 337
           (1976).

\end{description}
\end{document}